\documentclass[aps,prb,twocolumn,showpacs,superscriptaddress,longbibliography]{revtex4-1}
\usepackage{graphicx}  
\usepackage{dcolumn}   
\usepackage{bm}        
\usepackage{amssymb}   
\usepackage{multirow}
\usepackage{amsmath}
\usepackage{natbib}
\usepackage{xcolor}
\usepackage[unicode=true,colorlinks=true]{hyperref}
\newcommand{\affA}{Tyndall National Institute, Lee Maltings, Dyke Parade, Cork, Ireland}
\newcommand{\affB}{FOTON, Universit\'e Europ\'eenne de Bretagne, INSA-Rennes and CNRS, Rennes, France}
\newcommand{\affC}{Centre de Nanosciences et de Nanotechnologies, CNRS, Universit\'e Paris-Sud et Universit\'e Paris-Saclay, route de Nozay, 91460 Marcoussis, France}

\begin{document}

\title{Potential-inserted quantum well design for quantum cascade terahertz lasers}

\author{R.~Benchamekh}\affiliation{\affA}
\author{J.-M.~Jancu}
\affiliation{\affB}
\author{P.~Voisin}\affiliation{\affC}

\date{\today}
\begin{abstract}%
%
%
%
We report on a new design of terahertz quantum cascade laser based on a \textit{single}, potential-inserted  quantum well active region. The quantum well properties are engineered through single monolayer InAs inserts. The modeling is based on atomistic, spds* tight-binding calculations, and performances are compared to that of the classical three-well design. We obtain a 100\% increase of the oscillator strength per unit length, while maintaining a high, nearly temperature-independent contrast between phonon-induced relaxation times of the upper and lower lasing states. The improved performances are expected to allow THz lasing at room temperature.
 \end{abstract}

%
%


\maketitle   

Development of semiconductor-based terahertz (THz) lasers operating at a temperature reachable by thermo-electric cooling remains a major challenge in device physics.~\cite{Sirtori2013,Williams2007} The demonstration of THz quantum cascade lasers (QCL) was an important breakthrough,~\cite{Tredicucci2002} however the race to increase the operating range up to room temperature is still in progress. Different designs and engineering of both active region and collector/injector were proposed toward this goal. So far, the three quantum wells (QWs) design with tunnel- and phonon-assisted population inversion represent the state-of-the-art of THz QCLs,~\cite{Luo2007} with an actual operating temperature record of ~200K.~\cite{Fathololoumi:12}

Phonon-assisted population inversion is at the heart of THz QCL mechanism. To make this possible, one needs to engineer at least a three level system where the upper transition (laser transition) and the lower transition (relaxation transition) energies are respectively smaller and larger than the longitudinal optical (LO) phonon energy ($\hbar\omega_{LO}\sim$~36meV in GaAs). Indium and Aluminium isoelectronic monolayer insertions, initially proposed as a probe of the carrier wavefunction,~\cite{Marzin,Jancu1996} are an interesting tool to engineer the electronic states inside a QW. This was actually implemented to shorten the operating wavelength of mid infrared QCLs.~\cite{Wilson} Recently, it has been theoretically shown that a three level system suitable for THz applications can be obtained in a single unbiased GaAs/AlAs QW by including multiple In and Al monolayer insertions inside the well.~\cite{Raouafi20167} It was also shown that the envelope function approach (EFA), widely used in QCL modeling, fails to predict higher electron states energies in this particular case of very thin insertions, while the atomistic $spds^*$ tight-binding (TB) model gave essentially perfect agreement with existing spectroscopic data.

In this Letter we extend this work and report on a full design of a THz QCL based on a single QW active region containing a single In monolayer (ML) insertion. The main advantage is a large increase of lasing transition oscillator strength per unit length. Besides, in the proposed design the lower laser level has two distinct depopulation paths, either through a direct phonon assisted transition to a lower level in the active QW, or through a tunneling-assisted transition in the collector/injector QW. A large, nearly temperature-independent contrast between upper and lower states depopulation rates is obtained. These improvements in theoretical performances are expected to help reaching THz lasing at room temperature.

We begin with a presentation of the theoretical approach:  we use the extended-basis $spds^*$ TB model,~\cite{Jancu1998} with optimized TB  and strain parameters.~\cite{Nestoklon2016} We first define a supercell (containing 4$\times n$ atoms, where $n\sim400$ is the number of ML in the computed structure). Atomic positions are relaxed using Keating valence force field model.~\cite{Keating} Then the TB Hamiltonian is created including spin-orbit coupling, strain effects according to the strain model of Refs.~\citenum{Nestoklon2016} and~\citenum{Raouafi2016}, and the bias potential. Finally the Hamiltonian is diagonalized to determine the electronic state energies and wavefunctions.
\begin{figure}[b]
\includegraphics[width=0.99\columnwidth]{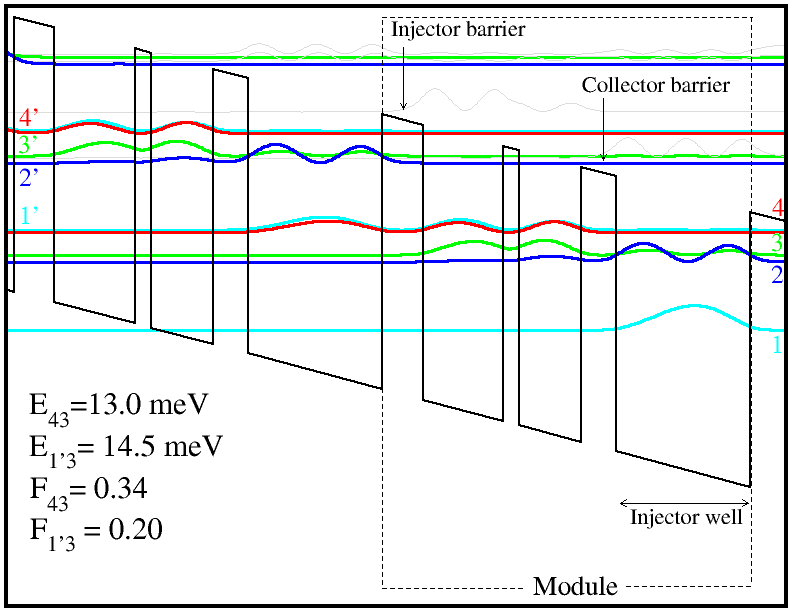}
\caption{Calculated conduction GaAs/Al$_{\text{0.15}}$Ga$_{\text{0.85}}$As band diagram under alignment bias (13~kV/cm). The $\bf{barrier}$ and well layer thicknesses are {\bf48}/96/{\bf20}/74/{\bf42}/161 \AA, starting from the injector barrier.}
\label{fig:1}
\end{figure}

Since the TB model was rarely used to model QCLs, we first validate our methodology by applying it to the well-documented 3-QW QCL design~\cite{Luo2007}. Figure~\ref{fig:1} shows the calculated conduction band diagram under alignment bias. Both the calculated transition energies and alignment field (13~kVcm$^{-1}$) are in excellent agreement with experimental results.~\cite{Luo2007} We note however that our result for the lasing transition oscillator strength ($F_{43}= 0.34$) is smaller than the value (0.51) reported in Ref.~\citenum{Luo2007}. This is likely due to the fact that we calculate dipole moments using the full spinorial wavefunctions while EFA adopts simplifying assumptions.
\begin{figure}
\includegraphics[width=0.7\columnwidth]{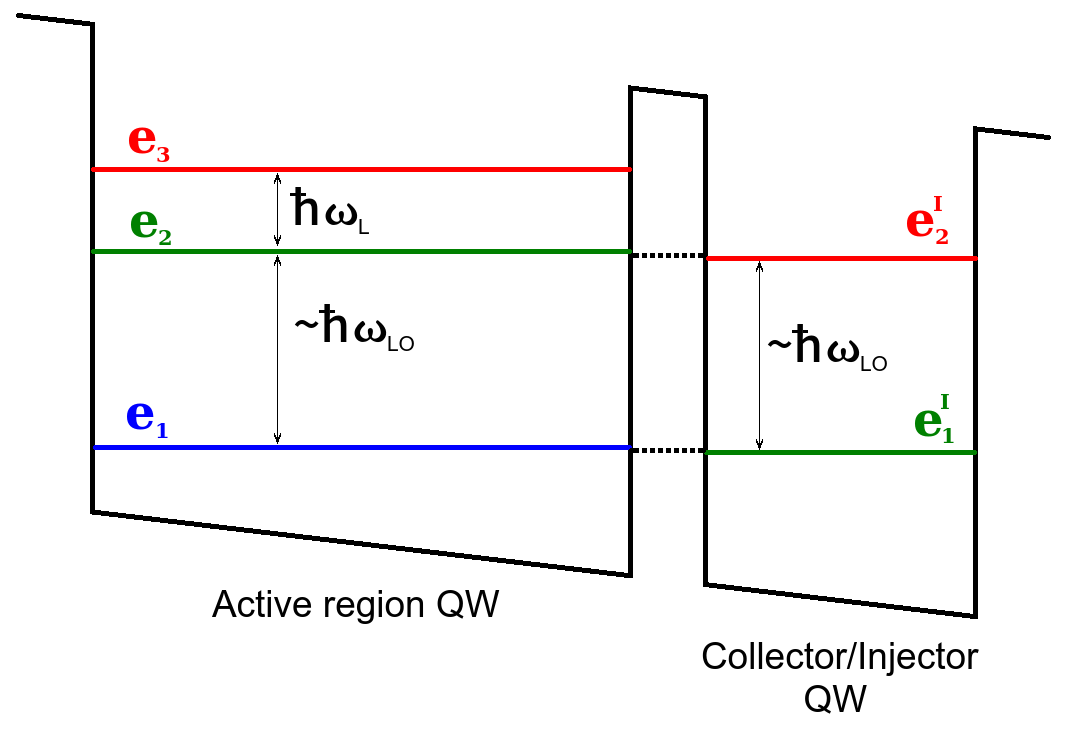}
\caption{Schematic representation of the targeted design.}\label{fig:2}
\end{figure}
\begin{figure}[b]
\includegraphics[width=0.99\columnwidth]{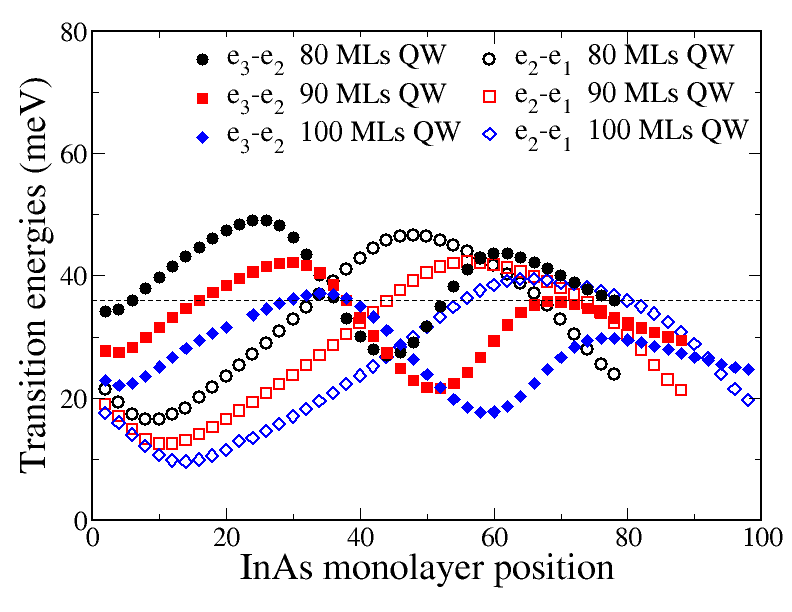}
\caption{Dependence of intersubband transition energies e$_2$-e$_1$ and e$_3$-e$_2$ on the In monolayer position for 80, 90 and 100 MLs thick GaAs/Al$_{0.2}$Ga$_{0.8}$As quantum wells, under a bias of 12 kV/cm. The dashed horizontal line corresponds to the LO phonon energy (36~meV).}\label{fig:3}
\end{figure}

Having checked our model against existing theoretical and experimental results, we now use it and design a new QCL structure. The idea is to create a single QW active region and a single QW collector/injector. The active region is engineered using In ML insertion in order to have the three lower electron state energies ($e_1$, $e_2$, $e_3$) fulfilling $e_3-e_2 \sim~15$ meV and $e_2-e_1 > \hbar\omega_{LO}$. It is obviously desirable that the energy separation between the two lower electron state ($e^I_1$ and $e^I_2$) of the collector/injector QW be equal to the energy separation between the two lower electron state in the active region ($e^I_2-e^I_1=e_2-e_1$). If the later condition is respected, at the alignement bias, the electrons in the lower laser state have two different depopulation paths (i) a phonon assisted transition to the state $e_1$ followed by a tunneling through the collector barrier to the collector state $e^I_1$ (ii) a tunneling to the collector state $e^I_2$ followed by a phonon assisted transition to $e^I_1$. A schematic representation of the targeted design is shown in Fig.~\ref{fig:2}.

The proposed double QW structure offers several tunable parameters: The QW widths, the barriers height, the barrier widths, the operating electric field, and the ML insertion positions. To reduce the number of parameters we arbitrarily fix the barrier composition to Al$_{0.2}$Ga$_{0.8}$As and the operating electric field to 12 kVcm$^{-1}$. These values may be slightly modified later to optimize the structure.
\begin{figure}[b]
\includegraphics[width=0.99\columnwidth]{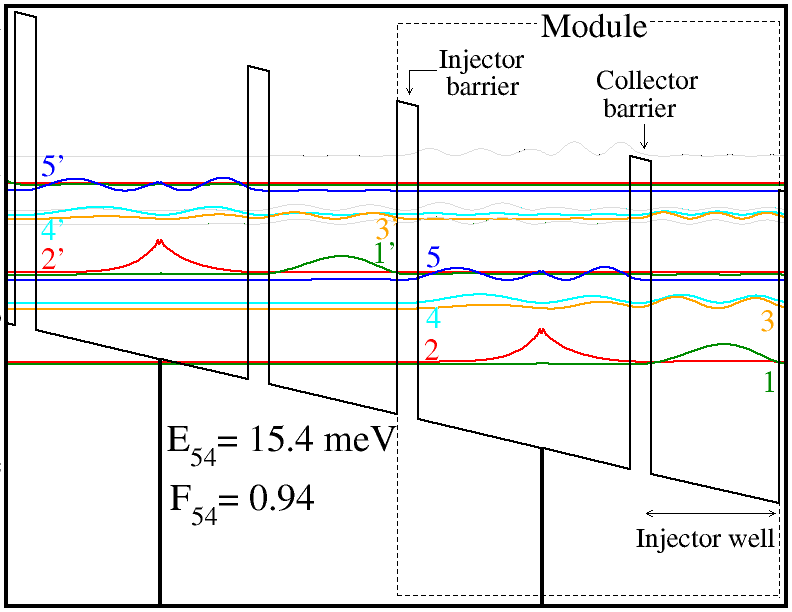}
\caption{Calculated conduction GaAs/Al$_{\text{0.2}}$Ga$_{\text{0.8}}$As band diagram under alignment bias (11.9~kV/cm). The $\bf{barrier}$ and well layer thicknesses are {\bf10}/58/\underline{1}/41/{\bf10}/60 MLs, starting from the injector barrier, and the InAs monolayer is underlined.}\label{fig:4}
\end{figure}

\begin{figure}[b]
\includegraphics[width=0.99\columnwidth]{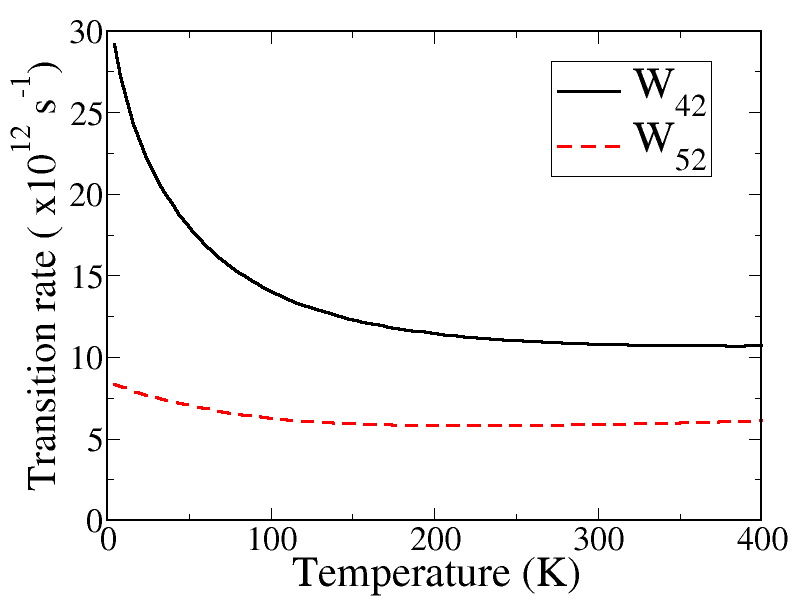}
\caption{Temperature dependence of the phonon emission rates of the 5$\rightarrow$2 and 4$\rightarrow$2 transitions in the QCL structure of Fig.~\ref{fig:4}.}\label{fig:5}
\end{figure}

\begin{table*}
\caption{Conduction state energies and lasing transition oscillator strength for different In concentration profile shapes and positions. $ z_{\text{In}}$ is in ML, the energies are given in meV with respect to the lowest conduction state in the module (e$_1$=0).}\label{tab:1}
\begin{tabular*}{2\columnwidth}{@{\extracolsep{\fill}}l|c|c|c|c|c|c|c|c|c|c}
\hline
\hline
$d_{\text{relax}}$ & $z_{\text{In}}$ & e$_2$ & e$_3$& e$_4$& e$_5$& e$_{1'}$& e$_5$-e$_3$ & F$_{53}$&e$_5$-e$_4$& F$_{54}$ \\
\hline
\multirow{4}{*}{1ML=2.83\AA} 
                               &  56 & 1.30 & 37.09 & 41.24 & 56.42 & 60.76 & 19.33 & 0.164 & 15.18 &  0.812 \\
                                                                      &  57 & 1.07 & 37.34 & 41.38 & 56.47 & 60.73 & 19.13 & 0.128 & 15.09 &  0.881 \\
                                                                      &  58 & 0.86 & 37.56 & 41.48 & 56.66 & 60.72 & 19.10 & 0.097 & 15.18 &  0.933 \\
                                                                      &  59 & 0.70 & 37.77 & 41.55 & 57.01 & 60.70 & 19.24 & 0.074 & 15.46 &  0.965 \\
\hline
\multirow{4}{*}{2ML=5.65\AA}
                               &  56 & 0.55 & 37.37 & 41.43 & 56.51 & 60.68 & 19.14 & 0.140 & 15.08 &  0.866 \\
                                                                      &  57 & 0.54 & 37.70 & 41.63 & 56.78 & 60.63 & 19.08 & 0.107 & 15.15 &  0.922 \\
                                                                      &  58 & 0.64 & 38.06 & 41.85 & 57.23 & 60.60 & 19.17 & 0.082 & 15.38 &  0.959 \\
                                                                      &  59 & 0.82 & 38.41 & 42.05 & 57.86 & 60.59 & 19.45 & 0.063 & 15.81 &  0.974 \\
\hline
\hline
\end{tabular*}
\end{table*}
We first study the effect of a single In ML insertion on the transition energies between the lower three conduction states in biased single QWs with different widths. We show on Fig.~\ref{fig:3} the dependence of e$_2$-e$_1$ and e$_3$-e$_2$ transition energies on the In ML insertion position for 80, 90 and 100 MLs thick GaAs/Al$_{0.2}$Ga$_{0.8}$As quantum wells, under a bias of 12 kV/cm. The ML insertion strongly modifies the transition energies, and 
a 3-level system with $e_3-e_2 < \hbar\omega_{LO}$ and $e_2-e_1 > \hbar\omega_{LO}$ can be obtained for various In ML positions. The minimum value of the transition energy $e_3-e_2$ is obtained for an In ML inserted away from the center of the well in the region with a lower electrostatic potential. This position corresponds to the maximum of the $e_1$ state wavefunction in the biased QW. Also the minimum value of the transition energy $e_3-e_2$ decreases when the QW thickness increases, as a direct consequence of the decrease of electron confinement energy. For the same reason the maximum of the transition energy $e_2-e_1$ decreases when the QW thickness increases. We finally conclude that a 100 ML thick QW with an In ML insertion at the 59$^{\text{th}}$ ML is a good starting point for the QCL structure design since  $e_2-e_1 = 38.0~\text{meV}$ and $e_3-e_2 = 17.6~\text{meV}$. 


Next we consider the injector/collector part of the QCL structure. We target a single QW injector/collector in which the transition energy between the lowest two electron state is close to the calculated e$_2$-e$_1$ in the active region QW ($\sim$38~meV). So we consider a single GaAs/Al$_{0.2}$Ga$_{0.8}$As QW under the same bias of 12 kV/cm, and vary its width. We find that the desired value of the transition energy is obtained for a QW width of about 60 MLs.

Finally we create the periodic QCL structure by coupling the two QWs engineered above. We choose the injector and collector barrier widths to be 10~MLs. This seems to be an appropriate value since, for a Al$_{0.2}$Ga$_{0.8}$As barrier, the resulting anticrossing energies are in the range 2~--~4~meV, which corresponds to tunneling times in the picosecond range. We finally tune the applied bias slightly to reach band alignment. Figure~\ref{fig:4} shows the calculated conduction band diagram for the designed structure. The laser transition (5$\rightarrow$4) has an energy of 15.4~meV (3.7~THz) and an oscillator strength per nanometer of 0.94~\cite{note1}, two times larger than in the structure of Ref.~\citenum{Luo2007} (c.f. Fig.~\ref{fig:1}). The lower laser state (labeled 4 on the figure) is aligned with second electron state in the collector, and is 39.7~meV above the lower electron state in the active region QW (labeled 2 on the figure). This design offers to the carriers in the lower laser state two different paths of depopulation: ({\it i}) a direct phonon-assisted transition to the lower state in the QW, or ({\it ii}) a tunneling through the collector barrier to the second electron state in the collector, followed by thermalization to ground state. Note that the transition 3$\rightarrow$1, in the injector well, has an energy equal to 37.6~meV, still larger than the LO-phonon energy. Hence the proposed design also has an undeniable advantage of offering two parallel phonon-assisted depopulation paths.


Although the proposed QCL structure has many advantages in terms of oscillator strength of the laser transition and phonon-assisted depopulation of the lower laser state, the population inversion would only be possible if the phonon-assisted transition 4$\rightarrow$2 is faster than the 5$\rightarrow$2. The modeling of lasing threshold, using advanced methods such as density matrix,~\cite{Dupont} non-equilibrium Green functions,~\cite{Lee,Kubis} or Monte Carlo techniques,~\cite{Callebaut} is far beyond the scope of this paper: here, we simply calculate the LO-phonon emission rates of the 5$\rightarrow$2 and 4$\rightarrow$2 transitions due to the Fr\"ohlich Hamiltonian, following the approach detailed in Ref.~\citenum{Smet96}. Figure~\ref{fig:5} shows the temperature dependence of the phonon-assisted 5$\rightarrow$2 and 4$\rightarrow$2 transition rates. At low temperature the 4$\rightarrow$2 is more than three time faster than the 5$\rightarrow$2 transition, and it is still two time faster at high temperature. These results prove that a population inversion between the levels 4 and 5 of the considered QCL structure is possible, and that the carriers lifetime in these two states have almost the same temperature dependencies, indicating a stronger stability against temperature raise. Note that the second depopulation path of level 4 is not included in this calculation, so actual situation will in fact be much better. A simple performance comparison between the here proposed design and the structure of Ref.~\cite{Luo2007} is possible, assuming similar tunneling extraction rates in both structures: At high temperature, when the tunneling extraction rates are drastically reduced, the direct phonon-assisted depopulation of the lower lasing state remains faster than the depopulation of the higher lasing state, indicating that our design may operate at a higher temperature.

Finally, we discuss the issue of design stability against fabrication defects. Indeed, In monolayer insertions have been realized by molecular beam epitaxy~\cite{Marzin,Jancu1996,Wilson}, and two main deviations from the ideal design may appear: The segregation of In atoms~\cite{Moison}, and the fluctuation of the ML position due to the long fabrication time. To model a more realistic In concentration profile, we replace the In ML in our design by an In segregation profile in the form: $ 1/d_{\text{relax}}Y(z-z_{\text{In}})(exp(-(z-z_{\text{In}}/d_{\text{relax}})) $, where $ z_{\text{In}}$ is the nominal insert position, $Y$ is the step function and $d_{\text{relax}} $ is a constant experimentally measured to be in the range 3--6 \AA  ~for InAs/GaAs ML. InGaAs alloys are modeled using the virtual crystal approximation model of Ref.~\citenum{Nestoklon2016b}. We calculate the conduction band diagram of the structure for two different values of $d_{\text{relax}}$ (1 and 2 ML) and for different positions of the In ML. Table~\ref{tab:1} summarizes the results. For the considered $d_{\text{relax}} $ and growth orientations, the band alignment is stable for a wide range of InAs ML position with no need to readjuste the operating bias. Also the lasing transition energy fluctuates by less then 1~meV, while its oscillator strength drops only by less than 25\% at worse.
We also considered the case where the tail of the exponential function is oriented in the opposite way (toward the injector rather than toward the collector), but obviously, results are not much different.

In summary, we report here a new design of THz QCLs based on a potential-inserted single QW active region and a single QW injector/collector. It offers significant advantages over the classical 3-well design in terms of transition oscillator strength and phonon-assisted depopulation of lower lasing state, and a limited penalty due to finite phonon-assisted depopulation of upper lasing transition. Generally speaking, the proposed design has many degrees of flexibility for ultimate tuning and optimization. We believe that this structure is a promising candidate to achieve room temperature THz QCLs. 

The authors thank Dr.  R. Colombelli for enlightening discussions.

\bibliography{Benchamekh_QCL}

\begin{thebibliography}{21}
\bibitem [{not()}]{note1}%
  \BibitemOpen
  \href@noop {} {}\bibinfo {note} {Note that the calculated transition energies
  for the single QW have been slightly modified due to the tunnel coupling with
  the injector/collector wells.}\BibitemShut {Stop}%
\end{thebibliography}
\bibliographystyle{apsrev4-1}
\end{document}